Relativistic five-quark equations and hybrid baryon spectroscopy.

Gerasyuta S.M. [1,2,3], Kochkin V.I. [1]

1. Department of Theoretical Physics, St. Petersburg State University, 198904, St. Petersburg, Russia.
2. Department of Physics, LTA, 194021 St. Petersburg, Russia.
3. Forschungzentrum Julich, Institut fur Kernphysik (Theorie)
   D-52425 Julich, Germany.

Abstract.

The relativistic five-quark equations are found in the framework of the dispersion relation technique. The behavior of the low-energy five-particle amplitude is determined by its leading singularities in the pair invariant masses. The solutions of these equations using the method based on the extraction leading singularities of the amplitudes are obtained. The mass spectra of nucleon and $\Delta$–isobar hybrid baryons are calculated. The calculations of hybrid baryon amplitudes estimate the contributions of four subamplitudes. The main contributions to the hybrid baryon amplitude are determined by the subamplitudes, which include the excited gluon states.

PACS numbers: 11.55 Fv, 12.39 Ki, 12.39 Mk, 12.40 Yx

## I. Introduction

The first calculation of light hybrid baryons in bag model was due to Barnes and Close [1], who derived the spectrum of nonstrange $qqqg$ states. The lightest hybrid baryon was predicted at about 1.6 GeV, which might possibly be identified with the Roper resonance. The calculation by Golowich, Haqq and Karl [2] gave a mass about 1.5 GeV for the lightest hybrid. The QCD sum rule calculations by Kisslinger and Li [3] estimated this mass about 1.5 GeV. A more recent review by Kisslinger [4] concluded that the Roper is a hybrid (meaning dominantly $qqqg$ state). Preliminary results for first $\frac{1}{2}^{+} N$ resonance in LGT have been reported by Sasaki [5]. But this approach used for heavier quarks, has not yet been carried out with high statistics for light quark masses.

The flux-tube model [6, 7] assumes that glue forms a dynamically excitable tube between quark and antiquark and that the lightest hybrids are states in which this flux-tube is spatially excited.

The determination of excited states in the hybrid baryons sector ($qqq$ + spatially excited flux-tube) is a rather complicated problem. Flux-tube model predictions for the lightest hybrid baryons were reported by Capstick and Page in 1999 [8]. They find that the lightest hybrid baryons are two each of $\frac{1}{2}^{+} N$ and $\frac{3}{2}^{+} N$, all a mass of 1.87 GeV. The flux-tube hybrid baryon multiplet contains the states $(\frac{1}{2}^{+} N)^2$, $(\frac{3}{2}^{+} N)^2$; $(\frac{1}{2}^{+} \Delta)$, $(\frac{3}{2}^{+} \Delta)$, $(\frac{5}{2}^{+} \Delta)$. The flux-tube $\Delta$ – isobar hybrids predicted to be degenerate, with a mass 2.08 GeV.

The implications of this paper [8] for searches of hybrid baryons, including various experimental study strategies such as strong decays, EM couplings and production amplitudes, were recently reviewed by Page [9].

In the recent paper [10] the relativistic generalization of the four-body Faddeev - Yakubovsky type equations are represented in the form of dispersion relations over the two-body subenergy. We investigated the relativistic scattering four-body amplitudes of the constituent quarks of two flavors (u, d). The poles of these amplitudes determine the masses of the lowest hybrid mesons. The constituent quark is color triplet and quark amplitudes obey the global color symmetry. We used the results of the bootstrap quark model [11] and introduced the $q\bar{q}$ – state in the color octet channel with $J^{PC} = 1^{--}$ and isospin I = 0. This bound state should be identified as constituent gluon. In our consideration we take into account the color octet state with $J^{PC} = 1^{--}$ and isospin I = 1, which determines with the constituent gluon the hybrid state. In addition, $q\bar{q}q\bar{q}$ states are also predicted.

We received the mixing hybrid and $q\bar{q}q\bar{q}$ states. This state was called the hybrid meson. The mass spectrum of lowest hybrid mesons with isospin I = 1 both exotic quantum numbers (non - $q\bar{q}$ ) $J^{PC} = 1^{-+}$, $0^{--}$ and ordinary quantum numbers $J^{PC} = 0^{++}$, $1^{++}$, $2^{++}$, $0^{-+}$, $1^{--}$ was calculated. The interesting result of this model is the calculation of hybrid meson amplitudes, which contain the contribution of two subamplitudes: four-quark amplitude and hybrid amplitude. One determined that the main contribution corresponds to the four-quark amplitude. The hybrid amplitude gives rise to only less 40 % of the hybrid meson contributions.

In the present paper the relativistic five-quark equations (like Faddeev – Yakubovsky equations) are constructed in the form of the dispersion relation over the two-body subenergy. The five-quark amplitudes for low-lying hybrid baryons are calculated under the condition that flavor SU(3) symmetry holds. It should be noted, that even in this approximation, the calculated masses of low-lying hybrid baryons (Tables 1, 2) agree well with experimental data [12] and with the results obtained in the flux-tube model [8].

In our relativistic quark model with four-fermion interaction the octet color $q\bar{q}$ bound state was found, which corresponds to the constituent gluon with mass $M_G = 0.67$ GeV [11]. This approach is similar to the large $N_c$ limit [13 - 15]. In diquark channel we have the diquark level with $J^P = 0^+$ and the mass $m_{ud} = 0.72$ GeV (in the color state $\bar{3}_c$). The diquark state with $J^P = 1^+$ in color state $\bar{3}_c$ also has the attractive interaction, but smaller than that of the diquark with $J^P = 0^+$, therefore there is only the correlation of quarks, not a bound state [11]. We can obtain the excited states of gluon (the excited quark-antiquark pair) for P-wave ($1^+, 2^+$) and D-wave ($2^-, 3^-$). In our model the lowest excited gluon states $P(0^+)$ and $D(1^-)$ - waves cannot be used to the construction of hybrid baryon bound states due to the small attractive interaction of quarks.

If we try to receive the hybrid baryons which are similar to paper [8], we can calculate the five lowest $N$ resonances:

two with $\qquad\qquad J^P = \frac{1}{2}^+ \qquad\qquad m = 1.870$ GeV

two with $\qquad\qquad J^P = \frac{3}{2}^+ \qquad\qquad m = 1.900$ GeV

one with $\qquad\qquad J^P = \frac{5}{2}^+ \qquad\qquad m = 1.973$ GeV

There we used the approach of paper [8].

The flux-tube model allows to receive only $J^P = \frac{1}{2}^+, \frac{3}{2}^+$ states. We calculated also $N$ state with $J^P = \frac{5}{2}^+$, but we have two contributions to the hybrid baryon amplitude: subamplitudes $qqqG$ and $qDG$, where $G$ is constituent gluon with $J^P = 1^-$ and the angular momentum L = 1 [8], $D$ is the diquark with $J^P = 0^+$.

If we use only first subamplitude, we receive the similar result to the calculations of paper [8]. In this case the state $N$ with $J^P = \frac{5}{2}^+$ is absent. In our consideration the main role in the construction of hybrid baryon amplitudes play the two subamplitudes: $qqqG$ and $qDG$ (Fig. 1). If we use the diquark with $J^P = 1^+$, we calculate the low-lying $\Delta$–isobar hybrid baryons. The mass spectra of nucleon and $\Delta$–isobar hybrid baryons are calculated using the excited states of gluon: P-wave ($1^+, 2^+$) and D-wave ($2^-, 3^-$).

The paper is organized as follows.

After this introduction, we represent the five-quark amplitudes of hybrid baryons (section 2).

In the section 3, we report our numerical results (Tables 1, 2) and the last section is devoted to our discussion and conclusion.

In the Appendix A we receive the relations, which allow to pass from the integration over the cosines of the angles to the integration over subenergies.

In the Appendix B we search the integration contours of functions $J_1$, $J_2$, $J_3$, which are determined by the interaction of the five quarks.

In the Appendix C we consider the determinant of the algebraic equations, which allow to calculate the mass spectra of hybrid baryons.

In the Appendix D the quark-quark and quark-antiquark vertex functions and phase spaces for the hybrid baryons are given respectively (Table 3, 4).

## II. Five-quark amplitudes of hybrid baryons.

We derived the relativistic five-quark equations in the framework of the dispersion relation technique. For the sake of simplicity one considers the case of the $SU(3)_f$ - symmetry, that the masses of all quarks are equal. We use only planar diagrams, the other diagrams due to the rules of $1/N_c$ expansion [13-15] are neglected. The correct equations for the amplitude are obtained at the account of all possible subamplitudes. It corresponds to the division complete system into subsystems from the smaller number of particles. Then one should represent five-particle amplitude as a sum of ten subamplitudes: $A = A_{12} + A_{13} + A_{14} + A_{15} + A_{23} + A_{24} + A_{25} + A_{34} + A_{35} + A_{45}$. In our case all particles are identical, therefore we need to consider only one group of diagrams and the amplitude corresponding to them, for example $A_{12}$. The set of diagrams associated with the amplitude $A_{12}$ can be further broken down into four groups corresponding to amplitudes $A_1(s, s_{1234}, s_{12}, s_{34})$, $A_2(s, s_{1234}, s_{25}, s_{34})$, $A_3(s, s_{1234}, s_{12}, s_{123})$, $A_4(s, s_{1234}, s_{25}, s_{125})$ (Fig. 1). The antiquark is determined by the arrow, the other lines correspond to the quarks.

The diagram equations for the amplitudes $A_1(s, s_{1234}, s_{12}, s_{34})$ $(qDG)$, $A_2(s, s_{1234}, s_{25}, s_{34})$ $(D\bar{q}D)$, $A_3(s, s_{1234}, s_{12}, s_{123})$ $(qqqG)$, $A_4(s, s_{1234}, s_{25}, s_{125})$ $(qq\bar{q}D)$ are taking into account the contribution of excited gluon states (P-wave $J^P = 1^+, 2^+$, D-wave $J^P = 2^-, 3^-$) and diquark states with $J^P = 0^+, 1^+$. The coefficients are determined by the permutation of quarks [16, 17].

In order to represent the subamplitudes $A_1(s, s_{1234}, s_{12}, s_{34})$, $A_2(s, s_{1234}, s_{25}, s_{34})$, $A_3(s, s_{1234}, s_{12}, s_{123})$ and $A_4(s, s_{1234}, s_{25}, s_{125})$ in form of the dispersion relation it is necessary to define the amplitudes of quark-quark and quark-antiquark interaction $b_n(s_{ik})$. The quark amplitudes $q\bar{q} \to q\bar{q}$ and $qq \to qq$ are calculated in the framework of the dispersion N/D method with the input four-fermion interaction with quantum numbers of the gluon [11]. We use the results of our relativistic quark model [11] and write down the pair quarks amplitude in the form:

$$b_n(s_{ik}) = \frac{G_n^2(s_{ik})}{1 - B_n(s_{ik})}, \qquad (1)$$

$$B_n(s_{ik}) = \int_{4m^2}^{\Lambda} \frac{ds_{ik}^{'}}{\pi} \frac{\rho_n(s_{ik}^{'})G_n^2(s_{ik}^{'})}{s_{ik}^{'} - s_{ik}}. \qquad (2)$$

Here $s_{ik}$ is the two-particle subenergy squared, $s_{ijk}$ corresponds to the energy squared of particles $i$, $j$, $k$, $s_{ijkl}$ is the four-particle subenergy squared and $s$ is the system total energy squared. $G_n(s_{ik})$ are the quark-quark and quark-antiquark vertex function (Table 3). $B_n(s_{ik})$, $\rho_n(s_{ik})$ are the Chew-Mandelstam function with cut – off $\Lambda$ [18] and the phase space respectively (Appendix D, Table 4). There n=1 corresponds to $qq$-pair with $J^P = 0^+$ in the $\bar{3}_c$ color state, n=2 describes $qq$-pair with $J^P = 1^+$ in the $\bar{3}_c$ color state and n=3 corresponds to $q\bar{q}$-pair in color state $8_c$ (excited constituent gluon) which includes the states with quantum numbers: $J^P = 1^+, 2^+$ (P-wave) and $J^P = 2^-, 3^-$ (D-wave).

In the case in question the interacting quarks do not produce bound state, therefore the integration in (3) - (6) is carried out from the threshold $4m^2$ to the cut-off $\Lambda$. The integral equation systems, corresponding to Fig. 1 (excited constituent gluon with $J^P = 1^+$ and diquark with $J^P = 0^+$), can be described as:

$$A_1(s, s_{1234}, s_{12}, s_{34}) = \frac{\lambda_1 B_3(s_{12}) B_1(s_{34})}{[1 - B_3(s_{12})][1 - B_1(s_{34})]} + 6\hat{J}_2(3,1) A_4(s, s_{1234}, s_{23}^{'}, s_{234}^{'}) + \\ + 2\hat{J}_2(3,1) A_3(s, s_{1234}, s_{13}^{'}, s_{134}^{'}) + 6\hat{J}_1(3) A_3(s, s_{1234}, s_{15}^{'}, s_{125}) + 2\hat{J}_1(3) A_4(s, s_{1234}, s_{25}^{'}, s_{125}) \qquad , \quad (3)$$

$$A_2(s, s_{1234}, s_{25}, s_{34}) = \frac{\lambda_2 B_1(s_{25}) B_1(s_{34})}{[1 - B_1(s_{25})][1 - B_1(s_{34})]} + \\ + 12\hat{J}_2(1,1) A_4(s, s_{1234}, s_{23}^{'}, s_{234}^{'}) + 6\hat{J}_1(1) A_3(s, s_{1234}, s_{15}^{'}, s_{125}) \qquad , \quad (4)$$

$$A_3(s, s_{1234}, s_{12}, s_{123}) = \frac{\lambda_3 B_3(s_{12})}{1 - B_3(s_{12})} + 4\hat{J}_3(3) A_1(s, s_{1234}, s_{13}^{'}, s_{24}^{'}), \tag{5}$$

$$A_4(s, s_{1234}, s_{25}, s_{125}) = \frac{\lambda_4 B_1(s_{25})}{1 - B_1(s_{25})} + 4\hat{J}_3(1) A_2(s, s_{1234}, s_{35}^{'}, s_{24}^{'}) + 2\hat{J}_3(1) A_1(s, s_{1234}, s_{35}^{'}, s_{21}^{'}), \tag{6}$$

were $\lambda_i$ are the current constants. We introduce the integral operators:

$$\hat{J}_1(l) = \frac{G_l(s_{12})}{[1 - B_l(s_{12})]} \int\limits_{4m^2}^{\Lambda} \frac{ds_{12}^{'}}{\pi} \frac{G_l(s_{12}^{'})\rho_l(s_{12}^{'})}{s_{12}^{'} - s_{12}} \int\limits_{-1}^{+1} \frac{dz_1}{2}, \tag{7}$$

$$\hat{J}_2(l, p) = \frac{G_l(s_{12}) G_p(s_{34})}{[1 - B_l(s_{12})][1 - B_p(s_{34})]} \times$$
$$\times \int\limits_{4m^2}^{\Lambda} \frac{ds_{12}^{'}}{\pi} \frac{G_l(s_{12}^{'})\rho_l(s_{12}^{'})}{s_{12}^{'} - s_{12}} \int\limits_{4m^2}^{\Lambda} \frac{ds_{34}^{'}}{\pi} \frac{G_p(s_{34}^{'})\rho_p(s_{34}^{'})}{s_{34}^{'} - s_{34}} \int\limits_{-1}^{+1} \frac{dz_3}{2} \int\limits_{-1}^{+1} \frac{dz_4}{2}, \tag{8}$$

$$\hat{J}_3(l) = \frac{G_l(s_{12}, \widetilde{\Lambda})}{1 - B_l(s_{12}, \widetilde{\Lambda})} \times$$
$$\times \frac{1}{4\pi} \int\limits_{4m^2}^{\widetilde{\Lambda}} \frac{ds_{12}^{'}}{\pi} \frac{G_l(s_{12}^{'}, \widetilde{\Lambda})\rho_l(s_{12}^{'})}{s_{12}^{'} - s_{12}} \int\limits_{-1}^{+1} \frac{dz_1}{2} \int\limits_{z_2^{-}}^{+1} dz \int\limits_{z_2^{-}}^{z_2^{+}} dz_2 \frac{1}{\sqrt{1 - z^2 - z_1^2 - z_2^2 + 2zz_1z_2}}, \tag{9}$$

were $l, p$ are equal 1 or 3. If we use the diquark with quantum number $J^P = 1^+$, $l, p$ are equal 2 or 3. There $m$ is a quark mass.

Hereafter we suggest that some unknown (not large) contribution from small distances which might be taken into account with the help of cut-off procedure. In the (7) – (9) we choose the "hard" cutting, but we can use also the "soft" cutting, for instance $G_n(s_{ik}) = G_n \exp\left(-(s_{ik} - 4m^2)^2 / \Lambda^2\right)$. It do not change essentially the calculated mass spectrum.

In the equations (7) and (9) $z_1$ is the cosine of the angle between the relative momentum of the particles 1 and 2 in the intermediate state and that of the particle 3 in the final state, which is taken in the c.m. of particles 1 and 2. In the equation (9) $z$ is the cosine of the angle between the momentu of the particles 3 and 4 in the final state, which is taken in the c.m. of particles 1 and 2. $z_2$ is the cosine of the angle between the relative momentum of particles 1 and 2 in the intermediate state and the momentum of the particle 4 in the final state, which is taken in the c.m. of particles 1 and 2. In the equation (8) we have defined that: $z_3$ is the cosine of the angle between relative momentum of particles 1 and 2 in the intermediate state and that of the relative momentum of particles 3 and 4 in the intermediate state, which is taken in the c.m. of particles 1 and 2. $z_4$ is the cosine of the angle between the relative momentum of the particles 3 and 4 in the intermediate state and that of the momentum of the particle 1 in the intermediate state which is taken in the c.m. of particles 3, 4.

Using the relation of Appendix A we can pass from the integration over the cosines of the angles to the integration over the subenergies.

Let us extract two-particle singularities in the amplitudes $A_1(s,s_{1234},s_{12},s_{34})$, $A_2(s,s_{1234},s_{25},s_{34})$, $A_3(s,s_{1234},s_{12},s_{123})$ and $A_4(s,s_{1234},s_{25},s_{125})$:

$$A_1(s,s_{1234},s_{12},s_{34}) = \frac{\alpha_1(s,s_{1234},s_{12},s_{34})B_3(s_{12})B_1(s_{34})}{[1-B_3(s_{12})][1-B_1(s_{34})]}. \tag{10}$$

$$A_2(s,s_{1234},s_{25},s_{34}) = \frac{\alpha_2(s,s_{1234},s_{25},s_{34})B_1(s_{25})B_1(s_{34})}{[1-B_1(s_{25})][1-B_1(s_{34})]}. \tag{11}$$

$$A_3(s,s_{1234},s_{12},s_{123}) = \frac{\alpha_3(s,s_{1234},s_{12},s_{123})B_3(s_{12})}{1-B_3(s_{12})}, \tag{12}$$

$$A_4(s,s_{1234},s_{25},s_{125}) = \frac{\alpha_4(s,s_{1234},s_{25},s_{125})B_1(s_{25})}{1-B_1(s_{25})}, \tag{13}$$

We do not extract three- and four-particle singularities, because they are weaker than two-particle and taking into account in the functions $\alpha$. The similar extraction is used when we consider the contribution of other diquark ($J^P = 1^+$).

We used the classification of singularities, which was proposed in paper [19] for the two and three particle singularities. The construction of approximate solution of the (3) - (6) is based on the extraction of the leading singularities of the amplitudes. The main singularities in $s_{ik} \approx 4m^2$ are from pair rescattering of the particles i and k. First of all there are threshold square-root singularities. Also possible are pole singularities which correspond to the bound states. The diagrams of Fig.1 apart from two-particle singularities have the triangular singularities, the singularities correspond to the interaction of four and five particles. Such classification allows us to search the approximate solution of (3) - (6) by taking into account some definite number of leading singularities and neglecting all the weaker ones. We consider the approximation, which corresponds to two-particle, triangle, four- and five-particle singularities. The functions $\alpha_1(s, s_{1234}, s_{12}, s_{34})$, $\alpha_2(s, s_{1234}, s_{25}, s_{34})$, $\alpha_3(s, s_{1234}, s_{12}, s_{123})$ and $\alpha_4(s, s_{1234}, s_{25}, s_{125})$ are the smooth functions of $s_{ik}$, $s_{ijk}$, $s_{ijkl}$, $s$ as compared with the singular part of the amplitudes, hence they can be expanded in a series in the singularity point and only the first term of this series should be employed further. Using this classification one define the reduced amplitudes $\alpha_1$, $\alpha_2$, $\alpha_3$, $\alpha_4$ as well as the B-functions in the middle point of the physical region of Dalitz-plot at the point $s_0$:

$$s_0^{ik} = s_0 = \frac{s + 15m^2}{10} \tag{14}$$

$$s_{123} = 3s_0 - 3m^2 , \ s_{1234} = 6s_0 - 8m^2$$

Such a choice of point $s_0$ allows to replace the integral equations (3) - (6) (Fig. 1) by the algebraic equations (15) - (18) respectively:

$$\alpha_1 = \lambda_1 + 6\alpha_4 J_2(3,1,1) + 2\alpha_3 J_2(3,1,3) + 6\alpha_3 J_1(3,3,1) + 2\alpha_4 J_1(3,1,1) , \tag{15}$$

$$\alpha_2 = \lambda_2 + 12\alpha_4 J_2(1,1,1) + 6\alpha_3 J_1(1,3,1) , \tag{16}$$

$$\alpha_3 = \lambda_3 + 4\alpha_1 J_3(3,3,1) , \tag{17}$$

$$\alpha_4 = \lambda_4 + 4\alpha_2 J_3(1,1,1) + 2\alpha_1 J_3(1,1,3) . \tag{18}$$

We use the functions $J_1(l,p)$, $J_2(l,p,r)$, $J_3(l,p,r)$ $(l,p,r = 1, 2, 3)$:

$$J_1(l,p) = \frac{G_l^2(s_0^{12}) B_p(s_0^{15})}{B_l(s_0^{12})} \int\limits_{4m^2}^{\Lambda} \frac{ds_{12}'}{\pi} \frac{\rho_l(s_{12}')}{s_{12}' - s_0^{12}} \int\limits_{-1}^{+1} \frac{dz_1}{2} \frac{1}{1 - B_p(s_{15}')} , \tag{19}$$

$$J_2(l,p,r) = \frac{G_l^2(s_0^{12}) G_p^2(s_0^{34}) B_r(s_0^{13})}{B_l(s_0^{12}) B_p(s_0^{34})} \times$$
$$\times \int\limits_{4m^2}^{\Lambda} \frac{ds_{12}'}{\pi} \frac{\rho_l(s_{12}')}{s_{12}' - s_0^{12}} \int\limits_{4m^2}^{\Lambda} \frac{ds_{34}'}{\pi} \frac{\rho_p(s_{34}')}{s_{34}' - s_0^{34}} \int\limits_{-1}^{+1} \frac{dz_3}{2} \int\limits_{-1}^{+1} \frac{dz_4}{2} \frac{1}{1 - B_r(s_{13}')} \tag{20}$$

$$J_3(l,p,r) = \frac{G_l^2(s_0^{12}, \widetilde{\Lambda}) B_p(s_0^{13}) B_r(s_0^{24})}{1 - B_l(s_0^{12}, \widetilde{\Lambda})} \frac{1 - B_l(s_0^{12})}{B_l(s_0^{12})} \times$$
$$\times \frac{1}{4\pi} \int\limits_{4m^2}^{\widetilde{\Lambda}} \frac{ds_{12}'}{\pi} \frac{\rho_l(s_{12}')}{s_{12}' - s_0^{12}} \int\limits_{-1}^{+1} \frac{dz_1}{2} \int\limits_{-1}^{+1} dz \int\limits_{z_2^-}^{z_2^+} dz_2 \frac{1}{\sqrt{1 - z^2 - z_1^2 - z_2^2 + 2zz_1z_2}} \frac{1}{[1 - B_p(s_{13}')][1 - B_r(s_{24}')]} \tag{21}$$

One obtained, that the other choice of point $s_0$ do not change essentially the contributions of $\alpha_1$, $\alpha_2$, $\alpha_3$ and $\alpha_4$, therefore we omit the indexes $s_0^{ik}$. That the vertex functions depend only slightly on energy it possibly to treat them as constants in our approximation and determine them in a way similar to that used in [20, 21].

The integration contours of functions $J_1$, $J_2$, $J_3$ are given in the Appendix B (Figs. 4, 5, 6). The equations, which are similar (15) – (18), correspond to other excited gluon states (graphic equations Figs. 2, 3) and are considered in the Appendix C. If we consider the mass spectrum of nucleon hybrid baryons, we describe the contributions of the excited constituent gluons with $J^P = 1^+, 2^+$ (P-wave), $J^P = 2^-, 3^-$ (D-wave) and the diquark with $J^P = 0^+$. In the case of the $\Delta$ - isobar hybrid baryons we must change quantum number of diquark ($J^P = 1^+$).

The solutions of the system of equations (Appendix C) are considered as:

$$\alpha_i(s) = F_i(s, \lambda_i) / D(s), \qquad (22)$$

where zeros of $D(s)$ determinants define the masses of bound states of hybrid baryons. $F_i(s, \lambda_i)$ are the functions of $s$ and $\lambda_i$. The functions $F_i(s, \lambda_i)$ determine the contributions of subamplitudes to the hybrid baryon amplitude.

III. Calculation results.

The poles of the reduced amplitudes $\alpha_1$, $\alpha_2$, $\alpha_3$, $\alpha_4$ correspond to the bound states and determine the masses of $N$ and $\Delta$ – isobar hybrid baryons. In the considered calculation the quark masses is not fixed. In order to fix anyhow $m$, we assume $m = 405\,MeV$ ($m \geq \frac{1}{5} m_{\frac{1}{2}^+}(1990)$). The model in question have two parameters: cut-off parameter $\Lambda_{0^+} = 22$ and gluon coupling constant $g = 0.2083$, which can be determined by mean of fixing of nucleon hybrid baryons $m_{\frac{1}{2}^+}(1710)$ and $m_{\frac{1}{2}^+}(1990)$. The calculation values of  low-lying nucleon hybrid baryons are shown in Table 1. In Table 2 we consider the mass spectrum of $\Delta$ – isobar hybrid baryons. We predict the degeneration of some states. One use the similar parameters to $N$ hybrid baryons, only the cut-off parameter is changed $\Lambda_{1^+} = 32.4$. It is fitted by $\Delta$ – isobar $m_{\frac{7}{2}^+}(1950)$. The calculated result was compared with the experimental data [12]. We did not determine the states with $J^P = \frac{1}{2}^+$ and the mass of Roper resonance [22 – 26]. The lowest state with $J^P = \frac{1}{2}^+$ has the mass $m = 1710$ MeV. The lowest $\Delta$ – isobar hybrid baryon with $J^P = \frac{3}{2}^+$ and the mass $m = 1600$ MeV is calculated. The calculations of hybrid baryons amplitudes estimate the contributions of four subamplitudes. The main contributions to the hybrid baryons amplitude are determined by the subamplitudes, which include the excited gluon states. The Table 1, 2 include the contribution of following subamplitudes: $A_1$ $(qDG)$, $A_2$ $(D\overline{q}D)$, $A_3$ $(qqqG)$, $A_4$ $(qq\overline{q}D)$. We obtained that the contributions of subamplitudes $A_1$, $A_3$ are large. The contributions of subamplitudes $A_2$, $A_4$ are small or absent. We calculate also the hybrid baryon mass using the approach Capstick and Page [8]. We used the $qqq$ color octet and spatially symmetric subsystem which combined with the angular momentum of the gluon ($J^P = 1^-$, S - wave). The other results (Table 1, 2) in the case of orbital excited gluons $J^P = 1^+, 2^+$ (P – wave) and $J^P = 2^-, 3^-$ (D – wave) are calculated. We used two type of diquarks $J^P = 0^+, 1^+$. The lowest excited gluon states ($J^P = 0^+$ P – wave and $J^P = 1^-$ D - wave) did not allow to obtain the bound states of hybrid baryons due to the small attractive interaction of quarks.

IV Conclusion.

In strongly bound system of light quarks such as the baryons consideration, where $p/m \approx 1$ the approximation of nonrelativistic kinematics and dynamics not justified.

In our relativistic five-quark consideration (Faddeev – Yakubovsky type approach) we received the masses of low-lying hybrid baryons. We used $SU(3)_f$ symmetry. The quark amplitudes obey the global color symmetry. The masses of the constituent quarks are equal 405 MeV. We consider scattering amplitudes of the constituent quarks. The poles of these amplitudes determine the masses of low-lying hybrid baryons. The lowest nucleon hybrid baryon have the mass 1710 MeV. We did not receive the Roper resonance with mass about 1.5 GeV. Maybe, the structure of the Roper resonance can be described as the three-particle system [22 – 26]. We calculated the lowest $\Delta$ – isobar hybrid state with $J^P = \frac{3}{2}^+$ and mass 1.6 GeV (Table 2). Obtained five-quark amplitude consist of four subamplitudes $qDG$, $D\overline{q}D$, $qqqG$, $qq\overline{q}D$, where $D$ and $G$ are the diquark state and excited constituent gluon state respectively.

Unlike mesons, all half-integral spin and parity quantum numbers are allowed in the baryon sector, so that experiments may not simple search for baryons with exotic quantum numbers is order to identify such hybrid states. Furthermore, no decay channels are a priori forbidden. These two facts make identification of a baryonic hybrid singularly difficult [27 – 29]. We believe that was made only the first step for the consideration of five-particle systems.

We manage with the quarks, diquarks and constituent gluon as with real particles. However, in the soft region the quark diagrams should be treated as spectral integrals over quark mass with the spectral density $\rho(m^2)$: the integration over quark mass in the amplitudes puts away the quark singularities and introduces the hadron ones. One can believe that the approximation:

$$\rho(m^2) \Rightarrow \delta(m^2 - m_q^2) \qquad (23)$$

could be possible for the low-lying hadrons (here $m_q$ is the "mass" constituent quark). We hope the approach given by (23) is sufficient good for the calculation of the low-lying hybrid baryons being carried out here.

Acknowledgments.

One of authors S.M. Gerasyuta is indebted to Institut fur Kernphysik Forschungzentrum Julich for the hospitality where a part of this work was completed. The authors would like to thank T. Barnes, D.I. Diakonov, A.M. Green, S. Krewald, N.N. Nikolaev, Yu.A. Simonov for useful discussions. This research was supported in part by Russian Ministry of Education, Program "Universities of Russia" under Contract № 01.20.00. 06448.

We can go over from integration with respect of the cosines of angles to integration with respect to the energy variables by using the relations:

$$s_{13}^{'} = 2m^2 + \frac{s_{123} - s_{12}^{'} - m^2}{2} + \frac{z_1}{2}\sqrt{\frac{s_{12}^{'} - 4m^2}{s_{12}^{'}}[(s_{123} - s_{12}^{'} - m^2)^2 - 4s_{12}^{'}m^2]} \tag{A1}$$

$$s_{24}^{'} = 2m^2 + \frac{s_{124} - s_{12}^{'} - m^2}{2} + \frac{z_1}{2}\sqrt{\frac{s_{12}^{'} - 4m^2}{s_{12}^{'}}[(s_{124} - s_{12}^{'} - m^2)^2 - 4s_{12}^{'}m^2]} \tag{A2}$$

$$z = \frac{2s_{12}^{'}(s_{1234} + s_{12}^{'} - s_{123} - s_{124}) - (s_{123} - s_{12}^{'} - m^2)(s_{124} - s_{12}^{'} - m^2)}{\sqrt{[(s_{123} - s_{12}^{'} - m^2)^2 - 4m^2 s_{12}^{'}][(s_{124} - s_{12}^{'} - m^2)^2 - 4m^2 s_{12}^{'}]}} \tag{A3}$$

$$s_{134}^{'} = m^2 + s_{34}^{'} + \frac{s_{1234} - s_{12}^{'} - s_{34}^{'}}{2} + \frac{z_3}{2}\sqrt{\frac{s_{12}^{'} - 4m^2}{s_{12}^{'}}[(s_{1234} - s_{12}^{'} - s_{34}^{'})^2 - 4s_{12}^{'}s_{34}^{'}]} \tag{A4}$$

$$s_{13}^{'} = 2m^2 + \frac{s_{134}^{'} - s_{34}^{'} - m^2}{2} + \frac{z_4}{2}\sqrt{\frac{s_{34}^{'} - 4m^2}{s_{34}^{'}}[(s_{134}^{'} - s_{34}^{'} - m^2)^2 - 4m^2 s_{34}^{'}]} \tag{A5}$$

The integration in consideration take on the physical region, where $-1 \leq z_i \leq 1$ ($i = 1, 2, 3, 4$). Then one can define the integration region on the invariant variables. Therefore for $s_{124}^{'}$ we have condition $0 \leq z^2 \leq 1$,

$$s_{124}^{'\pm} = s_{12}^{'} + m^2 + \frac{(s_{1234} - s_{123} - m^2)(s_{123} + s_{12}^{'} - m^2)}{2s_{123}} \pm$$
$$\pm \frac{1}{2s_{123}}\sqrt{[(s_{123} - s_{12}^{'} - m^2)^2 - 4m^2 s_{12}^{'}][(s_{1234} - s_{123} - m^2)^2 - 4m^2 s_{123}]} \tag{A6}$$

and the region of integration on $s_{12}^{'}$ in $J_3$:

$$\widetilde{\Lambda} = \begin{cases} \Lambda, \; if \; \Lambda \leq (\sqrt{s_{123}} + m)^2 \\ (\sqrt{s_{123}} + m)^2, \; if \; \Lambda > (\sqrt{s_{123}} + m)^2 \end{cases} \tag{A7}$$

The integration contour 1 (Fig. 4) corresponds to the connection $s_{123} < (\sqrt{s_{12}} - m)^2$, the contour 2 is defined by the connection $(\sqrt{s_{12}} - m)^2 < s_{123} < (\sqrt{s_{12}} + m)^2$. The point $s_{123} = (\sqrt{s_{12}} - m)^2$ is not singular, that the round of this point at $s_{123} + i\varepsilon$ and $s_{123} - i\varepsilon$ gives identical result. $s_{123} = (\sqrt{s_{12}} + m)^2$ is the singular point, but in our case the integration contour can not pass through this point that the region in consideration is situated below the production threshold of the four particles $s_{1234} < 16m^2$. The similar situation for the integration over $s_{13}$ in the function $J_3$ is occurred. But the difference consists of the given integration region that is conducted between the complex conjugate points (contour 2 Fig. 4). In Fig. 4, 5b, 6 the dotted lines define the square-root cut of the Chew-Mandelstam functions. They correspond to two-particles threshold and also three-particles threshold in Fig. 5a. The integration contour 1 (Fig. 5a) is determined by $s_{1234} < (\sqrt{s_{12}} - \sqrt{s_{34}})^2$, the contour 2 corresponds to the case $(\sqrt{s_{12}} - \sqrt{s_{34}})^2 < s_{1234} < (\sqrt{s_{12}} + \sqrt{s_{34}})^2$. $s_{1234} = (\sqrt{s_{12}} - \sqrt{s_{34}})^2$ is not singular point, that the round of this point at $s_{1234} + i\varepsilon$ and $s_{1234} - i\varepsilon$ gives identical results. The integration contour 1 (Fig. 5b) is determined by region $s_{1234} < (\sqrt{s_{12}} - \sqrt{s_{34}})^2$ and $s_{134} < (\sqrt{s_{34}} - m)^2$, the integration contour 2 corresponds to $s_{1234} < (\sqrt{s_{12}} - \sqrt{s_{34}})^2$ and $(\sqrt{s_{34}} - m)^2 \leq s_{134} < (\sqrt{s_{34}} + m)^2$. The contour 3 is defined by $(\sqrt{s_{12}} - \sqrt{s_{34}})^2 < s_{1234} < (\sqrt{s_{12}} + \sqrt{s_{34}})^2$. Here the singular point would be $s_{134} = (\sqrt{s_{34}} + m)^2$. But in our case this point is not achievable, if one has the condition $s_{1234} < 16m^2$. We have to consider the integration over $s_{24}$ in the function $J_3$. While $s_{124} < s_{12} + 5m^2$ the integration is conducted along the complex axis (the contour 1, Fig. 6). If we come to the point $s_{124} = s_{12} + 5m^2$, that the output into the square-root cut of Chew-Mandelstam function (contour 2, Fig. 6) is occurred. In this case the part of the integration contour in nonphysical region is situated and the integration contour along the real axis is conducted. The other part of integration contour corresponds to physical regions. This part of integration contour along the complex axis is conducted. The suggested calculation shows that the contribution of the integration over the nonphysical region is small [20, 21].

APPENDIX C

We considered the algebraic equations and determinants, which allow to calculate the poles of reduced amplitudes $\alpha_1$, $\alpha_2$, $\alpha_3$, $\alpha_4$ for the low-lying hybrid baryon. If we use the diquark with $J^P = 0^+$ ($l, p, r$ are equal 1 or 3), we can calculate the spectrum nucleon hybrid baryons. If we use the diquark with $J^P = 1^+$ ($l, p, r$ are equal 2 or 3), we can calculate the spectrum $\Delta$ - isobar hybrid baryons.

Fig. 1

$$\alpha_1 = \lambda_1 + 6\alpha_4 J_2(3,1,1) + 2\alpha_3 J_2(3,1,3) + 6\alpha_3 J_1(3,3,1) + 2\alpha_4 J_1(3,1,1)$$
$$\alpha_2 = \lambda_2 + 12\alpha_4 J_2(1,1,1) + 6\alpha_3 J_1(1,3,1)$$
$$\alpha_3 = \lambda_3 + 4\alpha_1 J_3(3,3,1) \qquad\qquad\qquad\qquad (C1)$$
$$\alpha_4 = \lambda_4 + 4\alpha_2 J_3(1,1,1) + 2\alpha_1 J_3(1,1,3)$$

$$D(s) = \{1 - 8J_3(3,3,1)[J_2(3,1,3) + 3J_1(3,3,1)]\}\{1 - 48J_2(1,1,1)J_3(1,1,1)\} - 4\{3J_2(3,1,1) + J_1(3,1,1)\}\times$$
$$\times\{J_3(1,1,3) + 48J_1(1,3,1)J_3(3,3,1)J_3(1,1,1)\}$$

Fig. 2

$$\alpha_1 = \lambda_1 + 6\alpha_3 J_2(3,1,1) + 2\alpha_2 J_2(3,1,3) + 6\alpha_2 J_1(3,3,1) + 2\alpha_3 J_1(3,1,1)$$
$$\alpha_2 = \lambda_2 + 4\alpha_1 J_3(3,3,1) \qquad\qquad\qquad\qquad (C2)$$
$$\alpha_3 = \lambda_3 + 2\alpha_1 J_3(1,1,3)$$
$$D(s) = 1 - 8J_3(3,3,1)\{J_2(3,1,3) + 3J_1(3,3,1)\} - 4J_3(1,1,3)\{3J_2(3,1,1) + J_1(3,1,1)\}$$

Fig. 3

$$\alpha_1 = \lambda_1 + 2\alpha_2 J_2(3,1,3) + 6\alpha_2 J_1(3,3,1)$$
$$\alpha_2 = \lambda_2 + 4\alpha_1 J_3(3,3,1) \qquad\qquad\qquad\qquad (C3)$$
$$D(s) = 1 - 8J_3(3,3,1)\{J_2(3,1,3) + 3J_1(3,3,1)\}$$

Functions $J_1(l,p,r)$, $J_2(l,p,r)$ and $J_3(l,p,r)$ correspond to (19) – (21), $l,p,r = 1, 2, 3$.

APPENDIX D

The vertex functions are shown in Table 3. The two-particle phase space for the equal quark masses is defined as:

$$\rho_n(s_{ik}, J^{PC}) = \left( \alpha(J^{PC}, n) \frac{s_{ik}}{4m^2} + \beta(J^{PC}, n) \right) \sqrt{\frac{s_{ik} - 4m^2}{s_{ik}}} \, ,$$

The coefficients $\alpha(J^{PC}, n)$ and $\beta(J^{PC}, n)$ are given in Table 4.

Table I. Low-lying nucleon hybrid baryon masses and contributions of subamplitudes $qDG$, $D\bar{q}D$, $qqqG$, $qq\bar{q}D$ to hybrid baryon amplitude in % (diquark with $J^P = 0^+$).

| Fig. № | Gluon $J^{PC}$ | $J^P$ | Mass, MeV | $qDG$ | $D\bar{q}D$ | $qqqG$ | $qq\bar{q}D$ |
|---|---|---|---|---|---|---|---|
| 1 | $1^{--}$ (S-wave) | $\frac{1}{2}^+$ | 1870 ( - ) | 36.83 | 8.93 | 43.45 | 10.79 |
| 2 | $1^{--}$ (S- wave) | $\frac{3}{2}^+$ | 1900 ( - ) | 40.29 | - | 51.30 | 8.41 |
| 3 | $1^{--}$ (S- wave) | $\frac{5}{2}^+$ | 1973 ( - ) | 42.23 | - | 57.77 | - |
| 1 | $1^{++}$ (P- wave) | $\frac{1}{2}^+$ | 1710 (1710) | 40.83 | 6.62 | 44.02 | 8.53 |
| 2 | $1^{++}$ (P- wave) | $\frac{3}{2}^+$ | 1730 (1720) | 43.30 | - | 49.52 | 7.18 |
| 2 | $2^{++}$ (P- wave) | $\frac{3}{2}^+$ | 1945 (1680) | 40.79 | - | 50.70 | 8.51 |
| 3 | $2^{++}$ (P- wave) | $\frac{5}{2}^+$ | 1990 (1990) | 40.03 | - | 59.97 | **-** |
| 2 | $2^{--}$ (D- wave) | $\frac{3}{2}^-$ | 1669 (1700) | 45.33 | - | 47.91 | 6.76 |
| 3 | $2^{--}$ (D- wave) | $\frac{5}{2}^-$ | 1754 (1675) | 44.81 | - | 55.19 | - |
| 3 | $3^{--}$ (D- wave) | $\frac{5}{2}^-, \frac{7}{2}^-$ | 1996 ( - ) | 39.82 | - | 60.18 | - |

Parameters of model: quark mass $m = 405$ MeV, cut-off parameter $\Lambda = 22$; gluon coupling constant $g = 0{,}2083$. Experimental mass values of nucleon hybrid baryons are given in parentheses [12].

Table II. Low-lying $\Delta$- isobar hybrid baryon masses and contributions of subamplitudes $qDG$, $D\bar{q}D$, $qqqG$, $qq\bar{q}D$ to hybrid baryon amplitude in % (diquark with $J^P = 1^+$).

| Fig. № | Gluon $J^{PC}$ | $J^P$ | Mass, MeV | $qDG$ | $D\bar{q}D$ | $qqqG$ | $qq\bar{q}D$ |
|---|---|---|---|---|---|---|---|
| 1 | $1^{++}$ (P- wave) | $\frac{1}{2}^+, \frac{3}{2}^+, \frac{5}{2}^+$ | 1600 (1600) | 54.81 | 2.36 | 38.02 | 4.81 |
| 1 | $2^{++}$ (P- wave) | $\frac{1}{2}^+, \frac{3}{2}^+, \frac{5}{2}^+$ | 1885 (1910) | 42.92 | 4.17 | 46.63 | 6.28 |
| 3 | $2^{++}$ (P- wave) | $\frac{7}{2}^+$ | 1950 (1950) | 44.88 | - | 55.12 | **-** |
| 1 | $2^{--}$ (D- wave) | $\frac{1}{2}^-, \frac{3}{2}^-, \frac{5}{2}^-$ | 1520 (1620) | 60.04 | 1.83 | 33.95 | 4.18 |
| 3 | $2^{--}$ (D- wave) | $\frac{7}{2}^-$ | 1698 ( - ) | 59.01 | - | 40.99 | - |
| 1 | $3^{--}$ (D- wave) | $\frac{3}{2}^-, \frac{5}{2}^-$ | 1900 (1700) | 44.13 | 4.15 | 45.50 | 6.22 |
| 3 | $3^{--}$ (D- wave) | $\frac{7}{2}^-, \frac{9}{2}^-$ | 1970 ( - ) | 46.61 | - | 53.39 | - |

Parameters of model: quark mass $m = 405$ MeV, cut-off parameter $\Lambda = 32.4$; gluon constant $g = 0{,}2083$. Experimental mass values of $\Delta$- isobar hybrid baryons are given in parentheses [12].

Table III. Vertex functions

| $J^{PC}$ | $G_n^2$ |
|----------|---------|
| $0^+$ (n=1) | $4g/3 - 8gm^2/(3s_{ik})$ |
| $1^+$ (n=2) | $2g/3$ |

$G_3^2 = 3g$, $g$ is the gluon coupling constant.

Table IV. Coefficient of Chew-Mandelstam functions for n = 3 (constituent gluon) and diquarks n = 1 ($J^P = 0^+$), n = 2 ($J^P = 1^+$).

| $J^{PC}$ | n | $\alpha(J^{PC}, n)$ | $\beta(J^{PC}, n)$ |
|----------|---|---------------------|--------------------|
| $1^{--}$ (S- wave) | 3 | 1/3 | 1/6 |
| $1^{++}$ (P- wave) | 3 | 1/2 | 0 |
| $2^{++}$ (P- wave) | 3 | 3/10 | 1/5 |
| $2^{--}$ (D- wave) | 3 | 4/7 | -1/14 |
| $3^{--}$ (D- wave) | 3 | 2/7 | 3/14 |
| $0^+$ (Diquark) | 1 | 1/2 | 0 |
| $1^+$ (Diquark) | 2 | 1/3 | 1/6 |

Figure captions.

Fig.1-3. Graphic representation of the equations for the five-quark subamplitudes $A_1(s, s_{1234}, s_{12}, s_{34})$ $(qDG)$, $A_2(s, s_{1234}, s_{25}, s_{34})$ $(D\bar{q}D)$, $A_3(s, s_{1234}, s_{12}, s_{123})$ $(qqqG)$, $A_4(s, s_{1234}, s_{25}, s_{125})$ $(qq\bar{q}D)$ using excited constituent gluons (S-wave $J^P = 1^-$, P-wave $J^P = 1^+, 2^+$, D-wave $J^P = 2^-, 3^-$) and diquarks with $J^P = 0^+, 1^+$.

Fig. 4. Contours of integration 1, 2 in the complex plane $s_{13}$ for the functions $J_1$, $J_3$.

Fig. 5. Contours of integration 1, 2, 3 in the complex plane $s_{134}$ (a) and $s_{13}$ (b) for the function $J_2$.

Fig. 6. Contours of integration 1, 2 in the complex plane $s_{24}$ for the function $J_3$.

Fig. 1

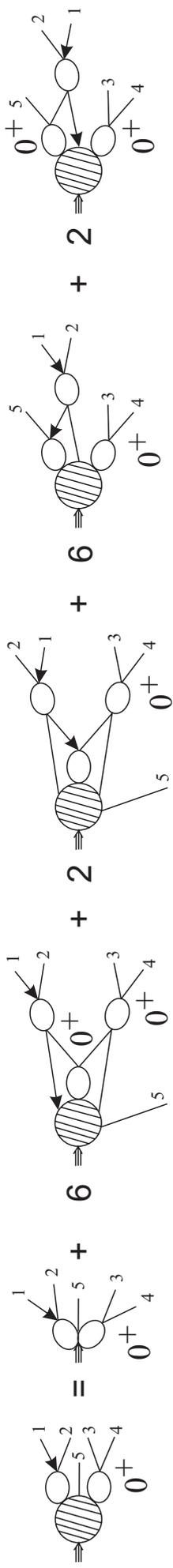

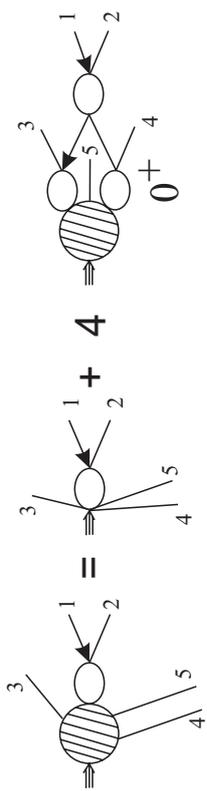

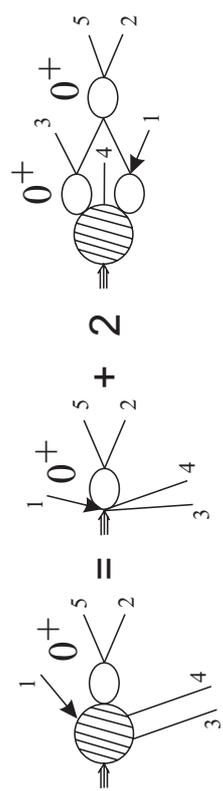

Fig. 2

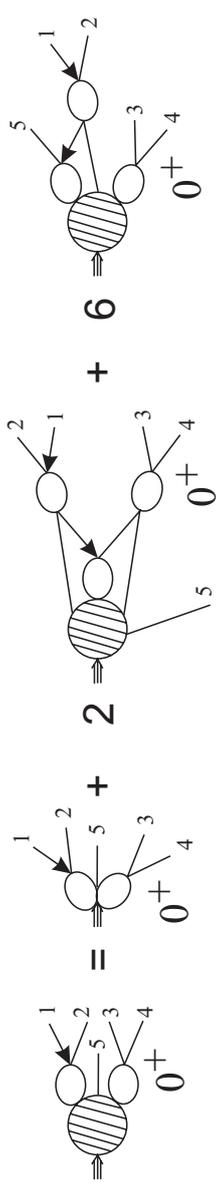

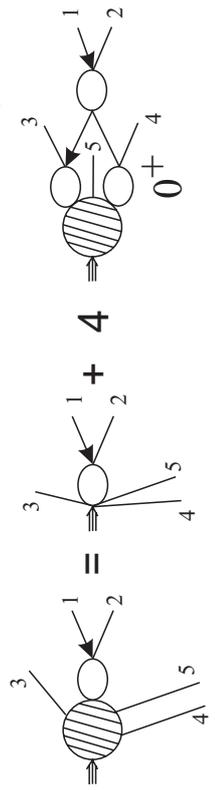

Fig. 3

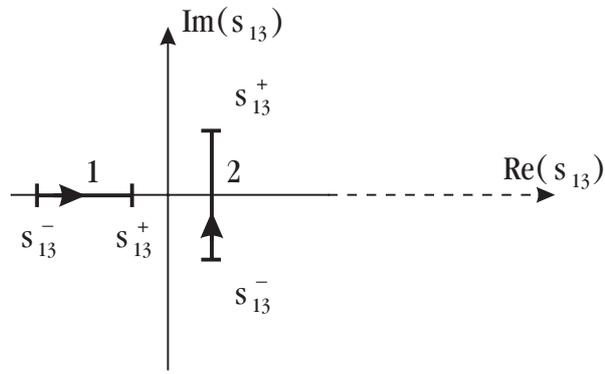

Fig. 4

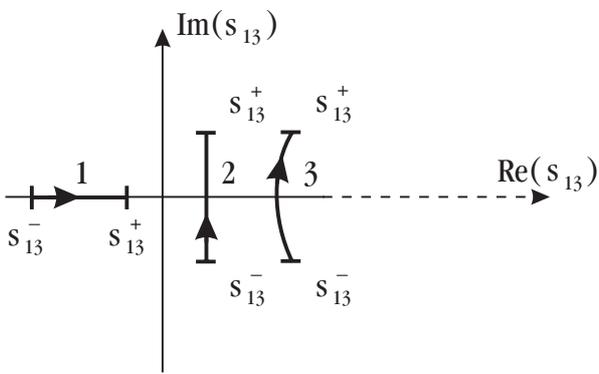

Fig. 5b

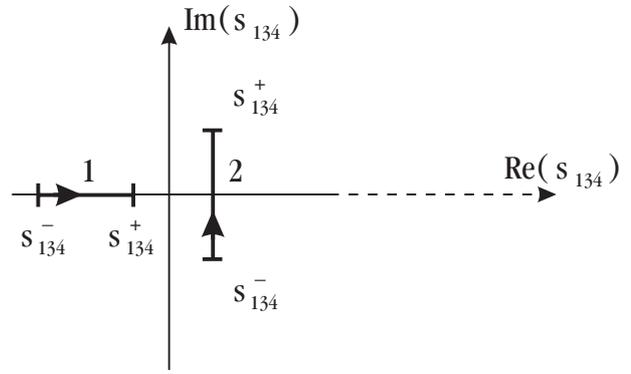

Fig. 5a

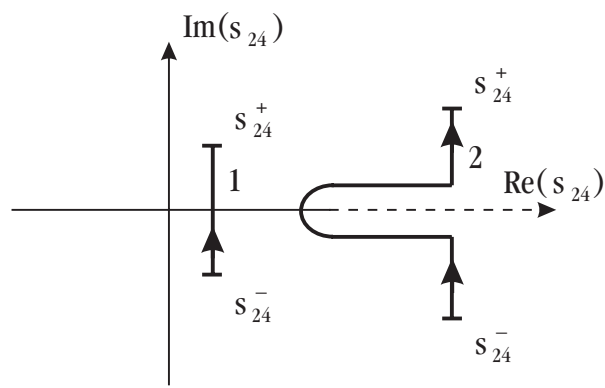

Fig. 6